\documentclass[smallextended]{svjour3}       
\smartqed  

\journalname{International Journal of Theoretical Physics}
\usepackage{graphics}
\usepackage{bm}
\usepackage{epigraph}
\usepackage{graphicx}

\newcommand{\beq}{\begin{equation}}
\newcommand{\eeq}{\end{equation}}
\newcommand{\beqa}{\begin{eqnarray}}
\newcommand{\eeqa}{\end{eqnarray}}
\def\nn{\nonumber\\}
\def\eq#1{(\ref{#1})}

\def\cd#1{\ensuremath{\nabla_{#1}}}          
\def\pd#1{\ensuremath{\partial_{#1}}}        

\def\st{space-time}

\def\mch{\scriptscriptstyle}

\def\Ch{\textstyle}

\def\text#1{{\rm #1}}

\def\lab#1{\label{#1}}

\begin{document}
\title{Empty  singularities in higher-dimensional Gravity}
\author{Ricardo E.\ Gamboa Sarav\'{\i} 
\thanks{\email{quique@fisica.unlp.edu.ar}}%
}                     
%
%
\institute{Departamento de
F\'\i sica, Facultad de Ciencias Exactas,\\ Universidad Nacional de
La Plata and IFLP,
CONICET.\\C.C. 67, 1900 La Plata, Argentina.}
\date{Received: date / Revised version: date}
%
%
\maketitle

\begin{abstract}
We study the exact solution of Einstein's field equations consisting  of a ($n+2$)-dimensional static and hyperplane symmetric  thick slice of matter, with constant and positive energy density $\rho$ and thickness $d$, surrounded by two different vacua.
We explicitly write down the pressure and the external gravitational
fields in terms of $\rho$ and  $d$, the pressure is
positive and bounded, presenting a maximum at an asymmetrical
position. And if $\sqrt{\rho}\,d$ is small enough, the dominant energy condition
is satisfied all over the spacetime.
We find that this solution presents many interesting features. In particular, it has an empty singular boundary in one of the vacua.
\end{abstract}


\section{Introduction}

We have recently  pointed out that solutions of  Einstein's field equations
presenting an empty (free of matter) repelling singular boundary where spacetime curvature diverges
can occur in four dimensions \cite{q1,q2,q3,q4}. These  singularities are not the sources of the fields, but rather, they arise owing to the attraction of distant matter.

The solution  described in  \cite{q1} is the
gravitational field  of a static and plane symmetric distribution of matter lying below $z=0$. Because of the symmetry required and depending on the properties of the matter, the  exterior solution can only turn out to be either  Rindler's flat spacetime or Taub's plane vacuum one \cite{taub}. Assuming the latter, which has a singular boundary at a finite height, it is showed that only vertical null geodesics just touch the singularity and bounce, whereas non-vertical null ones as well as massive particles bounce before getting to it. Furthermore, later on, we rigorously show that the Cauchy problem  for wave propagation in this vacuum is well-posed, if we only demand the waves to have finite energy, although no boundary condition is required \cite{GST}. And those waves completely reflect at the singularity. Due to these properties we also call this kind of singularities {\em white walls}.

A detailed study of  the inner solution  for a static and plane symmetric  relativistic perfect fluid obeying an equation of state such that $\rho$ and $p$ are proportional to each other is presented in \cite{q2}. In this case, it is not possible to match this solution to a vacuum one since the pressure does not vanish at any regular point. Nevertheless, it already exhibits by itself such a property because it is a semi-infinite spacetime where  pressure and density vanish at a singular boundary as well as at infinite.

An exhaustive  analysis of the properties of the internal
solution  for a static plane symmetric relativistic perfect incompressible fluid with positive density,  originally  found by  Taub in \cite{taub2}, is shown in \cite{q3}. We found that this solution finishes up down below
at an inner singularity at finite depth $d$. 
This solution essentially depends on only one parameter $\kappa$,  and depending on its value, it turns out to be gravitationally attractive, neutral or repulsive.

We analyze the $C^1$-matching of this inner solution and vacua in  \cite{q4}. In particular, we  consider  a  non-singular slice of the inner solution with
thickness $d$ $\left(0<d<\sqrt{\frac{\pi}{24\rho}}\,\right)$ surrounded by two
external vacua.   The solution turns out to be
attractive and remarkably asymmetric: the ``upper" solution is
Rindler's vacuum, whereas the ``lower" one is the singular part of
Taub's plane symmetric solution which finishes up
down below at a boundary  where spacetime
curvature diverges. We
explicitly write down the pressure and the external gravitational
fields in terms of $\rho$ and  $d$, the pressure is
positive and bounded presenting a maximum at an asymmetrical
position. And if $0<\sqrt{6\pi\rho}\,d<1.527\dots$, the dominant energy condition
is satisfied all over the spacetime. This exact and complete (matter matched to vacua) solution clearly shows how the
attraction of distant matter can shrink  \st\ in such a way
that it ends up at an  empty singular boundary.

Due to the plane symmetry  these solutions may look somewhat unnatural, nevertheless they
can give an idea of the qualitative features that
could arise in General Relativity,  and so, of possible properties
of realistic solutions.

\bigskip

The possibility that spacetime may have more than four dimensions is now a
standard assumption in High Energy Physics. Indeed,  most theories attempting to unify all the fundamental interactions and quantize gravity require a higher dimensional spacetime, where higher-dimensional gravity (HDG) plays a decisive roll. Therefore, among the reasons it should be interesting to study HDG we may mention that

\begin{itemize}

\item
String and M-theories contain gravity and requiere the existence of more than four \st\ dimensions, namely ten or eleven.

\item

The AdS/CFT correspondence \cite{M} states that the physics of gravity in a higher-dimensional spacetime, is equivalent to a certain quantum field theory which is defined on the boundary  (for a review see, for example, \cite{AGMOO}).

\item

According to world-brane models the observable universe could be $1+3$-surface  (the ``brane") embeded in a high-dimensional sapacetime (the ``bulk"), with Standard Model particles and fields trapped on the brane while gravity is free to acces the bulk (for a review see, for example, \cite{MK}).

\item
HDR is also interesting from a purely mathematical point of view. It is noteworthy that some known results of general relativity, turn out to be quite specific to four dimensions. For example, existence of  stable bound orbits or black hole no-hair theorem \cite{C}.

\end{itemize}

It is well known that Einstein's Gravity gets weaker as  the number of \st\  dimensions increases. For instance,  stable bound Keplerian orbits do not exist in \st s of more than four dimensions  \cite{Tang}, or also, the degree of compactification of spherical stars diminishes as the dimensionality of \st\  raises \cite{Cruz}.

On the other hand, HDG seems to be richer. Indeed, a much larger variety of asymptotically flat stationary black holes exists in spacetimes with extra dimensions. For instance, besides the Myer-Perry solution \cite{MP}, in five dimensions there exist black rings, black saturns, \dots (for a review see, for example, \cite{ER}).

Thus, it would be worthwhile   finding out whether  empty singular boundaries can also arise in higher-dimensional Gravity,  and the aim of this paper is to explore this issue.

In this work, we  show that such  singularities may occur  in \st s of $n+2$ dimensions ($n\ge2$)  by constructing an explicit solution.

The solution to be described is a ($n+2$)-dimensional static and hyperplane symmetric \st\ formed by a thick slice of matter, with constant and positive energy density $\rho$ and thickness $d$, surrounded by two vacua. 
It turns out that this solution shares all the main features of the one that was studied in \cite{q4} for the $n=2$ case. In particular, it has an empty singular boundary in one of the vacua.

For $n\geq2$, we adopt the convention in which the metric of the $(n+2)$-dimensional \st \
has signature $(-\ +\ \dots +\ +)$ and the  system of units in which
the speed of light $c=1$, and $G$ is the $(n+2)$-dimensional  Newton's gravitational
constant.

\section{The symmetry and Einstein's field equations}

We here consider  solutions  of Einstein's
equations corresponding to  a static and hyperplane symmetric
distribution of matter. That is, they must be
invariant under $n$ translations  and under  rotations in $n(n-1)/2$ planes. The matter we consider is  a perfect fluid with stress-energy tensor
$$ T_{ab}= (\rho+p)\,u_au_b+p\, g_{ab}\, ,$$ where $u^a$ is
the velocity of fluid elements.

Due to the required symmetry, as shown in \ref{apendice},   we can find coordinates such that
the line element can be written as
 \beq \lab {met} ds^2= - \mathcal{G}(z)^2\,dt^2+ e^{2V(z)}\left((dx^{\mch
1})^2+\dots+(dx^{\mch{n}})^2
\right)+dz^2\,.\eeq%

The non identically vanishing components of the Einstein tensor
are
\beqa \lab {gtt}   G_{tt}=-n\,\mathcal{G}^2 \left( V''+\,\frac{n+1}{2}\ V'^2\right),\\
\lab {gii}   G_{11}=\dots=G_{nn}= e^{2V} \left(\frac{\mathcal{G}''}{\mathcal{G}}+(n-1)\frac{\mathcal{G}'}{\mathcal{G}}\,V'+(n-1)V''+\frac{n(n-1)}{2}V'^2\right)\,,\\
\lab {gzz}   G_{zz}= n\ V' \left( \frac{\mathcal{G}'}{\mathcal{G}}+
\frac{n-1}{2} V' \right) ,
 \eeqa where a prime $(')$ denotes differentiation with respect to $z$.%

On the other hand, since the fluid must be static,
$u_a=(-\mathcal{G},0,\dots,0)$,  so \beq T_{ab}=
\text{diag}\left(\rho\, \mathcal{G}^{2},p\, e^{2V},\dots,p\,
e^{2V},p\right)\,, \eeq where $\rho$ and
 $p$  can depend only on  $z$.
Thus, Einstein's equations, i.e., $G_{ab}=4 \pi G \frac{n}{n-1}\ T_{ab}$ \footnote {We have set the normalization constant such that, in the Newtonian limit, they lead to $\nabla^2 \Phi=-\frac12 \nabla^2(1+g_{tt})= 4 \pi G\, \rho$.}, are
\beqa \lab {gtt1} V''+\frac{n+1}{2}\, V'^2= - \frac{4 \pi\, G}{n-1}\ {\rho}\,, \\%
\lab {gii1}
\frac{\mathcal{G}''}{\mathcal{G}}+(n-1)\frac{\mathcal{G}'} {\mathcal{G}}\,V'+(n-1)V''+\frac{n(n-1)}{2}V'^2
=\frac{4 \pi n\,G }{n-1}\ {p}\, ,\\
\lab {gzz1}   \frac{\mathcal{G}'}{\mathcal{G}}\,V' +
\frac{(n-1)}{2} V'^2  = \frac{4 \pi\, G}{n-1}\ {p}\,.
 \eeqa%

Moreover, $\cd a T^{ab}=0$ yields
\beq \lab{ppr} p' = -(\rho+p)\,\frac{\mathcal{G}'}{\mathcal{G}}\,.\eeq Of course, due to Bianchi's identities, equations
(\ref{gtt1})--(\ref{ppr}) are not
independent.

Regarding the mirror symmetry,  it can be  shown  that,  independently of the equation of state and the dimension of \st ,
the solution cannot have a ``plane" of symmetry in a region where $\rho(z)\neq0$
 and $p(z)\geq0$. In order to see this, let us assume that $z=z_s$ is that ``plane", then it
must  hold that $ \mathcal{G}'= V' =p' =\rho'=0$ at $z_s$, and so from
\eq {gzz1} we get that also $p(z_s)=0$. Now, by differentiating \eq{ppr} and using
\eq{gtt1} and \eq{gii1}, we obtain $p''(z_s)=-4\pi G\,\rho^2<0$.

\section{Solution with constant and positive density $\rho$}

In this section, we consider the solution for matter having constant energy density  $\rho>0$.  Proceeding analogously as in  \cite{q3} and \cite{q4}, where the four-dimensional case is considered,  we  find the solution
\beq\lab{p}p(z)  = \frac{C_p }{\mathcal{G}(z)}-\rho,  \eeq
\beq \lab{V} V(z)=\ln\left(
{C_1\,\sin{u}}\right)^{\frac{2}{n+1}},\eeq
and
\beqa     \mathcal{G}
\lab {G} = C_3\, \frac{\cos u}{\left(\sin u \right)^{\frac{n-1}{n+1}}}+\frac{C_p}{\rho}\ \frac{(n+1)}{(3n+1)}\  {\sin^2\! u}\,\, _2F_1\!\Bigl(1,\frac{n}{n+1};\frac{5n+3}{2(n+1)};\sin^2u \Bigr),\nn \eeqa %
where $u=\sqrt{2\pi G\,\frac{n+1}{n-1}\,\rho}\ z+C_2$, $_2F_1(a,b;c;z)$ is the Gauss hypergeometric function (see for example \cite{tablarusa}) and $C_p$, $C_1$, $C_2$ and $C_3$ are arbitrary constants.

Therefore,  the line element \eq{met} becomes%
 \beqa \lab{met1}
ds^2=  -  \mathcal{G}(z)^{2}\,
 dt^2 + \left({C_1\,\sin u}\right)^{\frac{4}{n+1}}\left((dx^{\mch
1})^2+\dots+(dx^{\mch{n}})^2 \right)+dz^2\,.
 \eeqa%
This solution is the higher dimensional generalization of the one
found by Taub  \cite{taub2}.

By contracting the Ricci tensor, we get
\beq \lab{R} R(z)=\frac{8\pi G}{n-1}\Bigl(\rho-(n+1)p(z)\Bigr)
\ .\eeq

On the other hand, we see that the metric has a \st\, curvature singularity when
$\sin u=0$, since straightforward computation of
the scalar quadratic in the
Riemann tensor yields%
\beqa\lab{RR}    R_{abcd}R^{abcd}=&4
\left(\frac{\mathcal{G}''^2}{\mathcal{G}^2}+n\frac{\mathcal{G}'^2}{\mathcal{G}^2}\,V'^2\right)
+4n\left(\,V''^2+2\,V''V'^2+\frac{n+1}{2}\,V'^4\right)\nn =&64\, \pi
^2\, G^2\, \rho ^2\,\frac{n^2}{n^2-1}\, \left(\frac{1}{\sin
^4u}+\frac{(n+1)^2}{n^2 (n-1)}\left(1+\frac{p}{\rho}\right)^2 \right.\nn&\left.
-\frac{4(n+1)}{n^2 (n-1)}\left(1+\frac{p}{\rho}\right)+\frac{2(n+2)}{n^2 (n-1)}\right),\eeqa so $
R_{abcd}R^{abcd}\rightarrow\infty$ when $\sin u\rightarrow0$.

\section{Vacuum limits}

From the solution found above, we can  obtain  vacuum ones as a
limit. In fact, when $C_p =0$, it is clear from \eq{p} that
$p(z)=-\rho$, and the solution \eq{met1} turns out to be a vacuum
solution with a cosmological constant $\Lambda= \frac{4 \pi n\, G}{n-1}\ {\rho}$
 \beqa \lab{metNH}   ds^2=  -
{\cos^2u}\;\left({\sin u}\right)^{-2\left(\frac{n-1}{n+1}\right)}  \, dt^2 +  \left({\sin u}\right)^{\frac{4}{n+1}}\left((dx^{\mch
1})^2+\dots+(dx^{\mch{n}})^2 \right)+ dz^2 \nn
-\infty<t<\infty,\quad-\infty<x^{\mch 1}<\infty,\quad\dots,\quad-\infty<x^{\mch n}<\infty,\quad0<u<\pi,\nn \eeqa%
where $u=\sqrt{\frac{(n+1)\Lambda}{2n}}\ z+C_2$.  We get from \eq{R} that it is
a space-time with constant scalar curvature $\frac{2(n+2)}{n}\Lambda$, and from
\eq{RR} we get that \beqa\lab{RR2}
R_{abcd}R^{abcd}={4}\left(\frac{n-1}{n+1}\right)\Lambda^2 \left(\frac{1}{\sin
^4u}+\frac{2(n+2)}{(n-1)n^2}\right) . \eeqa
This solution is the generalization of the one what was found for $n=2$ in   \cite{NH}.

Now, we  take the limit $\Lambda\rightarrow0$
($\rho\rightarrow0$). By setting $C_2=\pi-\frac{n-1}{g_{\mch T}}\sqrt{\frac{\Lambda}{2n(n+1)}}$ and an appropriate rescaling of the coordinates
$\left\{t,x^{\mch 1},\dots,x^{\mch n}\right\}$, we can readily see that when
$\Lambda\rightarrow0$, \eq {metNH} becomes \beqa \lab{taub}    ds^2=-
\left(1-\Ch{\frac{n+1}{n-1}}\,g_{\mch T}\,z\right)^{-2\left(\frac{n-1}{n+1}\right)}  \hspace{-1mm} dt^2 \hspace{6cm} \nn +  \left(1-\Ch{\frac{n+1}{n-1}}\,g_{\mch T}\,z\right)^{\frac{4}{n+1}}\hspace{-1mm}\left((dx^{\mch
1})^2+\dots+(dx^{\mch{n}})^2 \right)+ dz^2,\nn
-\infty<t<\infty,\;-\infty<x^{\mch 1}<\infty,\;\dots,\;-\infty<x^{\mch n}<\infty,\;0<1-\Ch{\frac{n+1}{n-1}}\,g_{\mch T}\,z<\infty \,, \nn  \eeqa%
where $g_{\mch T}$ is an arbitrary constant.  In \eq{taub}, the coordinates
have been chosen in such a way that the solution describes a homogeneous
gravitational field $g_{\mch T}$ pointing in the negative $z$-direction in
a neighborhood of  $z=0$. The metric \eq{taub} is the generalization  Taubs's
 vacuum plane solution \cite{L}. Notice that, in this case, \eq{RR2} becomes
\beqa\lab{RR3}
R_{abcd}R^{abcd}=\frac{16\ n^2 (n^2-1) g_{\mch T}^4}{(n-1)^4}\ \frac{1}{\left(1-\frac{n+1}{n-1}\,g_{\mch T}\,z\right)^4}\, , \eeqa
hence, this \st \ finishes at a singular boundary at $z=\frac{n-1}{(n+1)g_{\mch T}}$.

On the other hand, by setting
$C_2=\frac{1}{g_{\mch R }}\sqrt{\frac{(n+1)\Lambda}{2n}}+\frac{\pi}{2}$ and an appropriate
rescaling of the coordinate $t$, we can readily see that when
$\Lambda\rightarrow0$, \eq {metNH} becomes \beqa \lab{rindler}
ds^2=- (1+g_{\mch R}\,z)^{{2}}\,
 dt^2 +(dx^{\mch 1})^2+\dots+(dx^{\mch{n}})^2+dz^2,\hspace{2cm}\nn
-\infty<t<\infty,\quad-\infty<x^{\mch 1}<\infty,\quad\dots,\quad-\infty<x^{\mch n}<\infty,\quad-\frac{1}{g_{\mch R}}<z<\infty \,, \nn \eeqa%
where $g_{\mch R}$ is an arbitrary constant,  and the coordinates have been
chosen in such a way that the solution  also describes a homogeneous
gravitational field $g_{\mch R}$ pointing in the negative $z$-direction in
a neighborhood of  $z=0$. The metric \eq{rindler} is a $(n+2)$-dimensional
Rindler's flat \st.

Therefore, there are two very different vacua admitting the imposed symmetry, and both of them will be required to match to the inner solution.

\section{Properties of the interior solution}

Here we  confine
our attention to positive values of $\rho$ and $C_p \neq0$.
Without loss of generality, proceeding as in  \cite{q3} and \cite{q4} by an appropriate rescaling of the coordinates
$\left\{t,x^{\mch 1},\dots,x^{\mch n}\right\}$,   we can transform away the irrelevant parameters and see that the solution essentially depends on two parameters
$\rho$ and $\kappa$. The line element  \eq {met1} reads
 \beqa \lab{met2} ds^2=  -  \mathcal{G}(z)^{2}\,
 dt^2 + \left({\sin u}\right)^{\frac{4}{n+1}}\left((dx^{\mch
1})^2+\dots+(dx^{\mch{n}})^2 \right)+dz^2\,.\nn
-\infty<t<\infty,\quad-\infty<x^{\mch 1}<\infty,\quad\dots,\quad-\infty<x^{\mch n}<\infty,\quad  \nn 0<u=\sqrt{2\pi G\,\frac{n+1}{n-1}\,\rho}\ z+C_2\leq\pi/2\ , \eeqa%
where \beqa \lab{G2}
\mathcal{G}(z)=
\frac{\left(\kappa-\kappa_{\text{crit}}\right){\cos u}+ \,
_2F_1\!\Bigl(-\frac{1}{2},\frac{1-n}{2(n+1)};\frac{1}{2};\cos^2u
\Bigr)}{(\sin u )^{\frac{n-1}{n+1}}} \eeqa %
and \beq \lab{kapa}
\kappa_{\text{crit}}=\frac{\sqrt{\pi}\;{\Gamma\left(\frac{3n+1}{2(n+1)}\right)}} {{\Gamma\left(\frac{n}{n+1}\right)}}\ .\eeq
The pressure  \eq {p} becomes \beq\lab{p3}p(z) = \rho
\left(\frac{1}{\mathcal{G}(z)}-1\right). \eeq

\begin{figure}[b]
\begin{center}
\includegraphics[width=.45 \textwidth ]{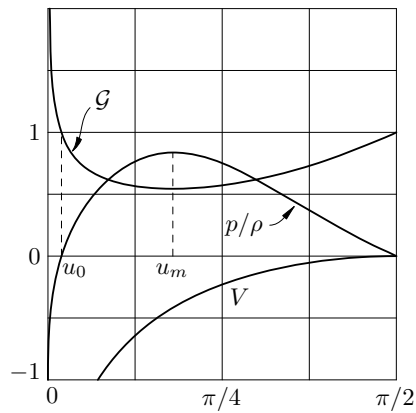}
\caption{\label{Gp}$\mathcal{G}(z)$, $V(z)$ and $p(z)/\rho$ as
functions of $u$, for $\kappa_{\text{crit}}>\kappa>\kappa_{dec}$. }\end{center}
\end{figure}

For $n=2$, in  \cite{q3} and \cite{q4} we present a detailed study of the properties of the function  $\mathcal{G}(z)$. Since for arbitrary $n$, the analysis completely parallels that one and its qualitative features do not depend on $n$, we  restrict ourselves to point out the relevant results. The interested reader can find them following the same steps  with slight modifications.

The  properties of  $\mathcal{G}$ drastically depend on the value of $\kappa$  (see Figs. 1 and 2 of Ref. \cite{q4}). From \eq{G2} we see that, independently of the value of $\kappa$,  $\mathcal{G}=1$ at $u=\pi/2$ and then $p$ vanishes  there. Moreover, we readily get from \eq{G2} that
 \beqa
\lab{G'1}\mathcal{G}'(z)|_{u=\pi/2}=\sqrt{\alpha}\, \left(\kappa_{\text{crit}}-\kappa\right)\,,
 \eeqa
where $\alpha={2\pi G\,\frac{n+1}{n-1}\,\rho}$, and so the parameter $\kappa$ essentially governs the derivative of $\mathcal{G}$ at $u=\pi/2$.

On the other hand, at $u=0$,  $\mathcal{G}$ diverges if $\kappa\neq0$,  whereas it vanishes at that point if $\kappa=0$. Therefore, at $u=0$,  $p=-\rho$ unless $\kappa=0$ in which case $p$ diverges.

If $\kappa\geq \kappa_{\text{crit}}$,  in the interval $0< u\leq\pi/2$,
$\mathcal{G}(z)$  decreases monotonically from $\infty$ to $1$. Consequently,   $p(z)$ is
negative when $0\leq u<\pi/2$ and it increases monotonically from
$-\rho$ to $0$  and it satisfies $|p|\le\rho$ all over the
space-time  (see Fig. 1(a) and Fig. 1(b) of Ref. \cite{q4}).

For $\kappa_{\text{crit}}>\kappa>0$,
there is one   (and only one) value  $u_{m}$ where  $\mathcal{G}(z)$ attains a local minimum. Hence,
there is one (and only one) value  $u_0$ ($0<u_0<\pi/2$) such that
$\mathcal{G}(z)|_{u=u_0}=\mathcal{G}(z)|_{u=\pi/2}=1$, and then
$\mathcal{G}(z)<1$ when $u_0<u<\pi/2$.
Since $\mathcal{G}(z)>0$, it is clear from \eq {p3} that $p(z)>0$
if $\mathcal{G}(z)<1$, and $p(z)$  reaches a maximum when
$\mathcal{G}(z)$ attains a minimum.
Hence, $p(z)$
grows from $-\rho$ to a maximum positive value when $u=u_{m}$
where it starts to decrease and vanishes at $u=\pi/2$. Thus,
$p(z)$ is negative when $0<u<u_0$ and positive when $u_0<u<\pi/2$
(see Fig. \ref{Gp}).

It can be readily seen
that, as $\kappa$ decreases from
$\kappa_{\text{crit}}$ to $0$, $u_m$ moves to the left and the
maximum value of $p(z)/\rho$ monotonically increases  from $0$ to
$\infty$. It can be shown that for
\beq  \lab{dec} \kappa=\kappa_{\text{dec}}:= \kappa_{\text{crit}}-\sqrt{\frac{2 n}{n+1}}
\left(\;_2F_1\!\left(-\frac{1}{2},\frac{1-n}{2 (n+1)};
\frac{1}{2};\frac{n+1}{2 n} \right)-\frac{1}{2}\left(\frac{n-1}{2n}\right)^{\frac{1-n}{2(n+1)}}\right)\,\eeq
this maximum value gets $1$, and then   $|p|\le\rho$ all over the
space-time for $\kappa\geq\kappa_{dec}$. Whereas, for
$ \kappa_{dec}>\kappa>0$, although the pressure is bounded everywhere, there is a region of space-time where the dominant energy condition is violated.

For $\kappa\leq0$, the pressure is unbounded (see Fig. 2 of Ref. \cite{q4}). If  $\kappa<0$, we see from \eq{RR} that a new curvature singularity appears where $p$ diverges.

\section{$C^1$ matching of solutions }

In this section, we  ensamble complete solutions by matching the interior solution and the vacuum ones. In order to avoid the appearance of undesirable superficial distribution of energy at the joint, we can readily see from \eq{gtt1} and \eq{gzz1} that the components of the metric tensor and their first derivatives must be continuous at the matching (hyper)surface.

Since for any value of
$\kappa$, the pressure vanishes at $u=\pi/2$, we can match the inner solution to a vacuum one there.

Now we see from \eq{V} that the interior $V'$ vanishes at $u=\pi/2$. On the other hand, we can readily see that  $V'$ does not vanish at any finite point for Taub's  vacuum \eq{taub}, whereas    $V'\equiv0$ for Ridler's vacuum \eq{rindler}. Therefore the inner solution can only be $C^1$-matched to the latter at $u=\pi/2$.

Since the field equations are invariant under $z$-translation, we
can choose to match the solutions at $z=0$ without losing
generality, so we select $C_2=\pi/2$. Thus, the inner singularity is located at $z=-\frac{\pi}{2 \sqrt{\alpha}}=-\sqrt{\frac{(n-1) \pi}{8 (n+1) G \rho}}$.

Equations \eq{rindler} and  \eq{met2} show that  at $z=0$ for both solutions, it holds that $g_{tt}=-1$, $g_{x^{\mch i}x^{\mch i}}=1$ and $\pd z g_{x^{\mch i}x^{\mch i}}=0$. Moreover  by using \eq{G'1} we see that the continuity of $\pd z g_{tt}$ at the  boundary yields \beq \lab{gR}g_{\mch R}=\sqrt{\alpha}\, \left(\kappa_{\text{crit}}-\kappa\right)\,,\eeq
which  relates  the  external gravitational field $g_{\mch R}$ with matter
density $\rho$ and $\kappa$.

The last equation shows that if
$\kappa>\kappa_{\text{crit}}$, $g$ is negative and  the  (hyper-)slab
turns out to be repulsive. If $\kappa=\kappa_{\text{crit}}$, it is
{\em gravitationally neutral}, and the exterior is one half of
Minkowski's space-time. If $\kappa<\kappa_{\text{crit}}$, it  is
attractive.

If $\kappa>0$,  the depth of the slab is $\sqrt{\frac{(n-1) \pi}{8 (n+1) G \rho}}$
independently of the value of
$\kappa$. In this case, the pressure is  finite anywhere, but it
is negative deep below and $p=-\rho$ at the inner singularity (see
Fig. 1 of Ref. \cite{q4}).  But, as
discussed above, only when
$\kappa\geq\kappa_{dec}$ is the condition $|p|\leq\rho$ everywhere
satisfied.

If $\kappa\leq0$,  the pressure   inside the slab is always
positive, and it diverges deep below at the inner singularity (see
Fig. 2 of Ref. \cite{q4}).

We have already seen that, in the case
$\kappa_{\text{crit}}>\kappa>0$, the pressure  also vanishes
inside the slab at the point where $u=u_0$ (see Fig. \ref{Gp}). Therefore, in this case, by matching the slice  of the interior solution $u_0\leq
u\leq\pi/2$ to a vacuum at $u_0$, we get an attractive slab surrounded by two vacua.

Clearly, the thickness of the slab  is given by
$d=\frac{(\pi/2-u_0)}{\sqrt{\alpha}}$, and  therefore $\left(0<d<\sqrt{\frac{(n-1) \pi}{8 (n+1) G \rho}}\ \right)$.
Now, since $\mathcal{G}(-d)=1$, from \eq{G2} we can write down
 $\kappa$ in terms of $d$  and $\rho$
\beqa \lab{kdd} \kappa=\kappa_{\text{crit}}+
\frac{{\left(\cos(\sqrt{\alpha}\, d)\right)^{\frac{n-1}{n+1}}}- \,
_2F_1\!\Bigl(-\frac{1}{2},\frac{1-n}{2(n+1)};\frac{1}{2};\sin^2(\sqrt{\alpha}\, d)
\Bigr)}{{\sin (\sqrt{\alpha}\, d)}}\,. \lab{kdch} \eeqa
From the last expression we can easily see   that $\kappa$ is a monotonically decreasing function of $d$. Therefore,  taking into account \eq{dec}, we find that, for every $n$, there exists a  constant $C_n$   such that for  $0<\sqrt{\rho}\,d<C_n$,  the dominant energy condition is satisfied anywhere.

In this case,  we see from \eq{V} that the interior $V'(-d)=\frac{2\sqrt{\alpha}}{n+1}\tan(\sqrt{\alpha}\, d)\neq0$. Therefore, the solution  can only  be matched to the Taub's vacuum
\beqa \lab{taub2}    ds^2=-
\left(1+\Ch{\frac{n+1}{n-1}}\,g_{\mch T}\,(z+d)\right)^{-2\left(\frac{n-1}{n+1}\right)}  \hspace{-1mm} dt^2\hspace{4cm}\nn +\,  C\,\left(1+\Ch{\frac{n+1}{n-1}}\,g_{\mch T}\,(z+d)\right)^{\frac{4}{n+1}}\hspace{-1mm}\left((dx^{\mch
1})^2+\dots+(dx^{\mch{n}})^2 \right)+ dz^2,\nn
-\infty<t<\infty,\;-\infty<x^{\mch 1}<\infty,\;\dots,\;-\infty<x^{\mch n}<\infty,\;-\Ch{\frac{n-1}{(n+1)\,g_{\mch T}}}-d<z<-d \,, \nn \eeqa
which  describes a homogeneous gravitational field $+g_{\mch T}$ in the
vertical  direction and finishes up at an empty singular boundary
at $z=-\Ch{\frac{n-1}{(n+1)\,g_{\mch T}}}-d$.

Since $ g_{tt}(-d)=-1$ and $
g_{ii}(-d)=C$, by comparing with \eq{met2},
we see that the continuity of the
metric components is assured if we set $C=\cos^{\frac{4}{n+1}}
(\sqrt{\alpha}\,d)$.  And, concerning the derivatives of
metric's components, from \eq{V} we get $\mathcal{G}'(-d )=-\frac{n-1}{2}V'(-d)$, and so their continuity requires
\beq \lab{gT} g_{\mch T}=\frac{(n-1)}{n+1}\,\sqrt{\alpha}\,\tan\left(\sqrt{\alpha}\,d\right)\,,\eeq
which  gives  the lower external gravitational field $g_{\mch T}$ in terms of matter
density $\rho$ and depth $d$.

Now, by eliminating $\kappa$ by means of \eq{kdd}, the solution
can be completely parameterized in terms of  $d$ and $\rho$,  \eq{G2}
becomes \beqa \lab{G4}   \mathcal{G}(z)=\frac{
\,_2F_1\!\Bigl(-\frac{1}{2},\frac{1-n}{2(n+1)};\frac{1}{2};\sin^2(\sqrt{\alpha}\,z)
\Bigr)}{{\cos^{\frac{n-1}{n+1}} (\sqrt{\alpha}\,z)}}
-\frac{{\cos^{\frac{n-1}{n+1}} (\sqrt{\alpha}\,d)}}{\cos^{\frac{n-1}{n+1}} (\sqrt{\alpha}\,z )}\ \ \frac{\sin
(\sqrt{\alpha}\,z)}{\sin
(\sqrt{\alpha}\,d)}\nn
 +\ \frac{{
_2F_1\!\Bigl(-\frac{1}{2},\frac{1-n}{2(n+1)};\frac{1}{2};\sin^2(\sqrt{\alpha}\,d)\Bigr)}}{\cos^{\frac{n-1}{n+1}}
(\sqrt{\alpha}\,z )}\ \ \frac{\sin
(\sqrt{\alpha}\,z)}{\sin
(\sqrt{\alpha}\,d)} \, ,\nn \eeqa %
the upper gravitational field \eq{gR}
\beq \lab{gr}
g_{\mch R}=\frac{\sqrt{\alpha}}{{\sin (\sqrt{\alpha}\,d)}}\left( { \,
_2F_1\!\Bigl(-\frac{1}{2},\frac{1-n}{2(n+1)};\frac{1}{2};\sin^2(\sqrt{\alpha}\,d)
\Bigr)}-{\cos^{\frac{n-1}{n+1}}\left(\sqrt{\alpha}\,d\right)}\right)\,,\eeq
and by using \eq{p3} and \eq{G4} we can write down the pressure.

This solution is remarkably asymmetric, not only because both
external gravitational fields are different, as one can  see
by comparing \eq{gT} and \eq{gr}, but
also because the nature of vacua is completely different: the
upper one is flat and semi-infinite, whereas the lower one is
curved and finishes up down below at an empty repelling boundary
where space-time curvature diverges.

It can be readily seen  that, in the upper vacuum, only vertically moving massless particles escape to infinite, whereas non-vertically moving massless ones as well as every massive particle, fall down after reaching a turning point. Consequently, in the lower one, only vertical null geodesics  touch the singularity and bounce, whereas non-vertical null ones as well as massive particles bounce before getting to it.

This exact solution clearly shows how the attraction of distant
matter can shrink the \st\ in such a way that it finishes at an
empty singular boundary, as pointed out in  \cite{q1} and \cite{q2}.

We can readily see from \eq{gT}, \eq{G4} and \eq{gr} that, in the Newtonian limit, {\em i.e.}, $\sqrt{\rho}\,d\ll1$, the mirror symmetry is restored.

For arbitrary $n\geq2$, we can also match    two interior solutions facing each
other as discussed in \cite{q3,q4}. Thus we consider two incompressible fluids joined at $u=\pi/2$ where
the pressure vanishes, the lower one having  density $\rho$ and
the upper one having  density $\rho'$. We can readily see that
$g_{tt}$, $g_{ii}$ and $\pd zg_{ii}$ are continuous at
the joint. Furthermore, from \eq{G'1}, we see that the continuity of
$\pd zg_{tt}$ requires \[
\sqrt{\rho}\left(\kappa_{\text{crit}}-\kappa\right)=-\sqrt{\rho'}
\left(\kappa_{\text{crit}}-\kappa'\right)\,.
 \]Therefore, the joining is only possible
between an attractive solution and a repulsive one, or between two
neutral ones.

Moreover, it is easy to see that we can also insert  a slice of arbitrary
thickness of the vacuum solution \eq{rindler} between two such solutions with  $(\kappa_{\text{crit}}-\kappa)$ of opposite sign, obtaining
a  relativistic plane ``gravitational capacitor".

\section{Concluding remarks }

In  order to find out whether empty singular boundaries can arise in higher dimensional Gravity, in this paper, we have constructed an exact and complete (matter matched to vacua) solution of Einstein's equations for arbitrary $n\geq2$.

This  space-time consists  of a ($n+2$)-dimensional static and hyperplane symmetric  thick slice of matter, with constant and positive energy density $\rho$ and thickness $d$, surrounded by two different vacua.

The solution turns out to be
remarkably asymmetric because the nature of both
vacua is completely different: the ``upper" one is flat and
semi-infinite, whereas the ``lower" one is curved and finishes up
down below at an empty repelling boundary where space-time
curvature diverges.

The pressure is positive and bounded,
presenting a maximum at an asymmetrical position between  the
boundaries. We explicitly wrote down the pressure  and the
external gravitational fields in terms of $\rho$ and  $d$. We showed
that if $\sqrt{\rho}\,d$ is small enough, the dominant energy
condition is satisfied all over the space-time.

This exact solution clearly show how the attraction of distant matter
can shrink the \st\ in such a way that it finishes at an empty
singular boundary, as pointed out in  \cite{q1} and \cite{q2}.

Therefore, in spite of the weakening of gravity with the number of dimensions of the space-time \cite{Tang,Cruz}, our solution clearly shows that it is still strong enough to generate  empty  singular boundaries  for arbitrary $n\geq2$.

\begin{appendix}
\section{Killing vectors  and adapted coordinates}\label{apendice}

We want to find  coordinates adapted  to  a  hyperplane symmetric
distribution of matter. That is, \st\  must be
invariant under $n$ translations  and under  rotations in $n(n-1)/2$ hyperplanes.

More precisely, a $n+2$ dimensional \st\, will be said to be
$n$-dimensional Euclidean homogenous if it admits the $r=n(n+1)/2$
parameter group of isometries of the $n$-dimensional Euclidean space ISO$(n)$.

Since the spacetime admits $n$  mutually conmuting independent
motions, we can choose coordinates $x^i,z,t$ so that the
corresponding Killing vectors are $\xi_{(i)}=\pd{i}$
($i=1,\dots,n$), and so \beq \xi_{(i)}^k =
\delta^k_i\ \ \ \ \ \ \
\left(\begin{array}{c}k=t,1,\dots,n,z\\i=1,\dots,n\end{array}\right)\,.\eeq
The equations of Killing, \beq \xi^k \pd {k}g_{ij}+ g_{kj}\ \pd
{i}\xi^k + g_{ik}\ \pd {j} \xi^k= 0\,, \label{killing}\eeq
corresponding to these vectors become \beq \label{aa} \pd {k}
g_{ij}= 0\ \ \ \ \ \ \ \ \ \ \
\left(\begin{array}{c}i,j=t,1,\dots,n,z\\k=1,\dots,n\end{array}\right)\,.\eeq
Hence all the components of the metric tensor depend only on the
coordinates $t$  and $z$ and the metric is unaltered by the finite
transformation \beq x^i \rightarrow x^i + a^i\ \ \ \ \ \ \
(\text{for}\ i=1,\dots ,n)\,.\eeq We can take for the remaining
$n(n-1)/2$ motions, the generators $\xi_{(ij)}= x^i \pd j - x^j \pd
i $ ($i<j\ \text{an d}\ i,j=1,\dots,n$), so \beq \label{ab}   \pd l
\xi_{(ij)}^k =\delta^i_l \delta^k_j-\delta^j_l \delta^k_i\ \ \ \ \
\ \  \left(\begin{array}{c}k,l=t,1,\dots,n,z\\i<j\ \text{and}\
i,j=1,\dots,n\end{array}\right)\,.\eeq Taking into account (\ref{aa})
and (\ref{ab}) from the equations of Killing (\ref{killing}), we
get \beq     g_{jm} \delta^i_l -g_{im} \delta^j_l+g_{lj}
\delta^i_m-g_{li} \delta^j_m= 0\ \ \ \ \ \ \
\left(\begin{array}{c}m,l=t,1,\dots,n,z\\i<j\ \text{and}\
i,j=1,\dots,n\end{array}\right)\,.\eeq From the last equation we
readily find \beq    g_{ii}=g_{jj} \ \ \ \text{and}\ \ \
g_{ij}=g_{it}=g_{iz}=0 \ \ \ \ \ \ \ (\ i\neq j\ \text{and}\
i,j=1,\dots,n)\,.\eeq Furthermore, we can take the bidimensional
metric of the $V_2$ spaces $x^i=$ constant ($i=1,\dots ,n$) in the
conformal flat form. Hence, the most general metric admitting this group of isometries  may be written as
 \beq  ds^2= - e^{2 U(z,
t)}\left(dt^2-dz^2\right)+ e^{2V(z, t)}\left((dx^{\mch
1})^2+\dots+(dx^{\mch{n}})^2
\right)\,.\eeq%

If, in addition, we impose staticity, $U$ and $V$ must be time independent, and the change of variable $\int e^{U(z)}dz \rightarrow z$ brings  the line element to the form \eq{met}.

\end{appendix}

\end{document}